\documentclass[bib]{ba}

\usepackage{bayes}
\usepackage{amsmath,amsfonts,amsthm,amssymb,latexsym}
\usepackage{natbib}

\newenvironment{kvothe}{\small\begin{flushright}}{\end{flushright}\normalsize\vglue -0.2truecm}
\newcommand\dropcap\noindent

\begin{document}

\inserttype{article}
\author{Chopin, N., Gelman, A., Mengersen, K.L., \&~Robert, C.P.}{%
  {\sc Nicolas Chopin}\\{\em ENSAE, and CREST, Paris}\\{\sf nicolas.chopin@ensae.dauphine.fr}\\
  {\sc Andrew Gelman}\\{\em Department of Statistics and Department of Political Science, Columbia University}\\ 
  {\sf gelman@stat.columbia.edu}\\
  {\sc Kerrie L. Mengersen}\\{\em Mathematical Sciences, Queensland University of Technology}\\
  {\sf k.mengersen@qut.edu.au}\\
  {\sc Christian P.~Robert}\\{\em Universit\'e Paris-Dauphine, CEREMADE, IUF, and CREST, Paris}\\{\sf xian@ceremade.dauphine.fr}
}       
\title[In praise of the referee]{In praise of the referee}
\date{17 May 2012}
\maketitle

\begin{abstract} There has been a lively debate in many fields, including statistics and related applied fields
such as psychology and biomedical research, on possible reforms of the scholarly publishing system.  Currently,
referees contribute so much to improve scientific papers, both directly through constructive criticism and
indirectly through the threat of rejection.  We discuss ways in which new approaches to journal publication
could continue to make use of the valuable efforts of peer reviewers.  \end{abstract}


\section{Introduction}

\begin{kvothe}
{\em ``Damn referees, I'll miss them less than anybody."} --- Abe Lemons
\end{kvothe}

Scholars have long been dissatisfied with aspects of the peer review system, but now there is an increasing
sense that something drastic should and could be done about it.   In the present article we discuss some recent
reform proposals given our own experiences publishing in theoretical and applied statistics.

Scholarly publishing has many problems, a lot of degrees of freedom, and various dimensions on which
one could measure success.  Put these together and you have the potential for unending opportunities for
change, with each new idea promising improvement in some aspect of the system.  Thus it is no surprise that
many different proposals have arisen for reforming the peer-review system, with each new plan sounding
completely reasonable but in contradiction to proposals by others.  We bring this up not as a way to cynically
dismiss innovation in this area but rather to explain how it is possible that so many completely different
ideas can seem to make so much sense.  It is a bit like reform of primary education:  some people recommend
back-to-basics while others recommend modern student-centred learning.  Either approach could make sense,
given the many different ways that such ideas can be implemented and the multiplicity of potential outcome
measures.

As statisticians, it would be most appropriate for us to evaluate reform proposals by analysing existing data,
systematically gathering new information, or at the very least proposing a plan for sampling, measurement, and
causal inference.  Unfortunately, as in much of our professional lives, we do not practice what we preach.
Instead of systematic data collection and analysis, we offer the same sort of mix of opinion and personal
experience that you might get from any other participant in the system.  Thus in this article we speak not as
statisticians but merely as a particular set of academic researchers with our own particular perspective.

\section{A peer-review system in flux}

\begin{kvothe}
{\em ``Off with the ref!"}
\end{kvothe}
Scientific and scholarly publishing has for many years been centred on peer-reviewed journals, where the
authors of published articles are responsible for their correctness, while editors and referees vouch for the
novelty and importance of the work.  Each field has a set of journals with a rough partial ordering (top
theoretical and applied journals, top sub-field journals, second-tier, and third-tier publications).

Widely-acknowledged problems with the current system include inefficiency for authors (time spent preparing and
revising a single article for multiple journals), waste of reviewers' efforts (long referee reports that are
read by only two people---the author and the journal editor), a breakdown of the sorting process (so that it is
no longer sufficient for scholars to keep up with a field by reading three or four top journals), the exclusion
of ``gated'' research from people without access to a major university library, and, most importantly, a
profusion of unreplicated or unreplicable claims even in the highest-prestige outlets.

For instance, Wasserman's (\citeyear{wasserman:2012}) remarks that ``we are using a refereeing system that is
almost 350 years old. If we used the same printing methods as we did in 1665 it would be considered
laughable.''  He describes the refereeing process as``'noisy, time consuming and arbitrary,'' that it ``limits
dissemination'' and that provides an ``illusion'' of quality control.  He likens the process to a
``priesthood'' or ``guild'' and advocates its replacement by a ``marketplace'' of ideas. 

Concerns with the peer-reviewed publication system are not new; they are, for example, integral to the relative
attribution of credit to Newton versus Leibniz for the invention of differential calculus in the 17th century
\citep{hall:1980}.

Proposals for reform typically choose among the following options:  (1) removing journals' roles as copyright
holders and gatekeepers by setting up incentives for researchers to post their articles freely on the web, (2)
formalising a post-publication peer-review process so that referee reports are open and available for all to
read, and (3) putting more of the burden of proof of replicability on published work by requiring data-based
articles to come with full replication materials. 


Each of these steps has been taken already, to some extent.  Personal websites and servers such as arXiv
(physics and mathematics) and SSRN (social science) are widely used for posting of unreviewed preprints.  You
can't come in off the street and publish at arXiv or SSRN but it is not difficult for a researcher to establish
the connections necessary to post there.  Post-publication peer review exists in some journals and, more
effectively, in an informal network of scientific blogs.  The goal of ensuring replicability is tougher, but
some journals (for example, the {\em Quarterly Journal of Political Science\/}) do require a full suite of replication
materials before allowing any empirical article to be published.



Thus, proposed reforms typically involve taking some aspect of the current system and pushing it further.  Here
are three recent suggestions:
\begin{enumerate}
\item Theoretical statistician Larry Wasserman (2012) calls for ``a world without referees'':
\begin{quotation}
\noindent
"We should be disseminating our research as widely as possible. Instead, we let two or three referees stand in
between our work and the rest of our field. (\dots)  We should think about our field like a marketplace of
ideas. Everyone should be free to put their ideas out there. There is no need for referees. Good ideas will get
recognised, used and cited. Bad ideas will be ignored."
\end{quotation}
\item Cognitive psychologist Nikolaus Kriegeskorte (2009, 2011) proposes ``open post-publication peer-review'':
\begin{quotation}
\noindent
"Any scientist can instantly publish a peer review on any published paper. The scientist will submit the review
to a public repository. (\dots) The repository will link each paper to all its reviews, such that that readers
are automatically presented with the evaluative meta-information. In addition, the repository allows anyone to
rank papers according to a personal objective function computed on the basis of the public reviews and their
numerical quality ratings."
\end{quotation}
\item Political scientist Brendan Nyhan (2012), following some ideas that have become popular in medical
research, recommends that data-collection protocols be published ahead of time, with the commitment to publish
the eventual results:
\begin{quotation}
\noindent
"In the case of experimental data, a better practice would be for journals to accept articles before the study
was conducted. The article should be written up to the point of the results section, which would then be
populated using a pre-specified analysis plan submitted by the author. The journal would then allow for
post-hoc analysis and interpretation by the author that would be labelled as such and distinguished from the
previously submitted material. By offering such an option, journals would create a positive incentive for
preregistration that would avoid file drawer bias. More published articles would have null findings (at least
5\%!), but that's how science is supposed to work."
\end{quotation}
\end{enumerate}
All three of these proposals are appealing, compelling, and radical---and go in different directions, with the
statistician wanting to eliminate referees, the psychologist recommending reviews but in a different structure,
and the political scientist proposing a more stringent system of pre-publication quality control.

Our goal is not to evaluate these particular proposals but rather (a) to consider the relevance of these ideas
for the fields of applied and computational statistics, and (b) to emphasise the value of the referee system
and to focus attention on how to not lose its benefits in this time of change..

This note summarises our reflections on the worth of the current and admittedly imperfect referee system, while
making constructive proposals towards its improvement, rather than its abolition.   We attempt to find a middle
ground between what we have now and various proposed reforms. In our opinion, the debate is as much about
ethics as it is about science, namely how to work out a system of dissemination in which papers are evaluated
on the basis of their scientific worth, rather than on the paper's conformity with existing norms (a problem
with the traditional system of peer review), its potential popular impact (an issue with proposed open
alternatives), the author's reputation or networks, or the reviewer's own long-term plans. 

Based on our own experiences, we argue that in this era of data explosion, the referee system remains
preferable to the frightening morass of an uncontrolled accumulation of self-published documents.

It can be instructive to compare a concern at hand with a parallel situation. Where else do we find this
practice of refereeing? In sport, of course. So we will make use of the cries heard during a soccer match as an
apt introduction to some of our points throughout the note.

\section{Background}

\begin{kvothe}
{\em ``Whadya call that, ref?!"}
\end{kvothe}

Each field or sub-field brings its own perspective on publishing.  For a mathematician or theoretical
statistician such as Wasserman, what's important in a publication is the idea.  Mathematical ideas can be
evaluated openly and, in principle, by anyone. From the other direction, Nyhan focuses on the {\em difficulty}
of replicating empirical results, especially given the selection problem that positive rather then negative
findings tend to get published.  As applied statisticians, we see the merits of both approaches, depending on
what we are working on.

At the same time that mathematicians are moving to deregulate the academic publication system, many
experimental scientists are pushing toward more formal registries. Beyond their direct benefit in
replicability, such reforms involve incentives for better behaviour on the part of researchers.  If you know
ahead of time that you will have to supply details of your design, methods, data and computer code, you will be
motivated to keep better records and clearer codes the first time, which in turn leads to a positive feedback
in which later analyses are improved by iterating on existing material.  This is the argument made by O'Rourke
and Detsky (\citeyear{orourkedetsky:1989}).

Publication patterns also vary among academic fields.  Some of the best mathematicians and economists work
alone or in small collaborations and publish papers after they have been honed by workshop and seminar
presentations, while, at the other extreme, leading physicists, biologists, and electrical engineers supervise
laboratories producing dozens of publications a year. In the first case, one could argue against an extra
refereeing stage, however biases in the workshop process also need to be ironed out by this anonymous
refereeing step.

Turning now to the analogy, we could ask the same question: why do we have referees in sport?  There is a
myriad of backyard and self-organised games around the world, but a neutral referee is demanded in major games.
(Of course, each side may also bring their own umpires or referees, which again has potential parallels with
current and possible future practices in scientific refereeing, but we will not pursue this here.) It is argued
that such a referee can maintain, and ensure, a high quality, fair and fast game for the benefit of both
players and spectators. Surely we ask the same of a scientific referee: assurance of quality---in terms of the
merit, originality and substantive contribution of the scientific content; fairness---in terms of equitable
treatment for all authors; consistency---in terms of reasonable, useful feedback to authors; and timeliness---a
fast turnaround of reviews. These are the very qualities that Wasserman (2012) laments are lacking in the
current process.

\section{Some horror stories with happy endings}

\begin{kvothe}
{\em ``He didn't touch him, ref!"}
\end{kvothe}

The previous section seems to make the usual argument that refereeing is a necessary evil. We believe on the
contrary that it is a necessary good. Yes, certain referees are annoying, or even aggressive or too dismissive
about one's work. Of course, like others, we can tell horror stories about referees completely missing the
point or even being outright dishonest. In the past decades, we indeed have all accumulated good and bad
experiences about the (human and hence imperfect!) refereeing process.  As authors of dozens or hundreds of
peer-reviewed publications, we have however benefited immensely from the unpaid labour of referees (while at the
same time having spent countless hours serving as referees, associate editors, and sometimes editors). 

At times, we've been annoyed at having to jump through hoops---and some of us prefer writing books and blogs,
where in both cases out our audience is the reader not the referee---but more often than not the suggestions
are helpful.  For example, Gelman's (\citeyear{gelman:2006}) most successful article of the past decade was his
paper on prior distributions for hierarchical variance parameters.  This article originated as an example in an
appendix of a book co-authored by Gelman ({\em Bayesian Data Analysis}),  and then was solicited as an article
by the editor of the then-new journal {\em Bayesian Analysis}.  The referees were brutal (maybe unfairly so,
considering the positive reaction the paper had had in the years since publication) and the paper could only be
published in the journal in the form of a discussion of another article on a related topic. However, as a
consequence of this revision process, Gelman was motivated to add a whole new section that made the research
much more general and interesting.  It is thanks to the referees that the author put in the work to make the
paper what it was. A similar experience occurred with the paper of Gelman and Rubin (1992).

Another extreme example experienced by Chopin is that of a referee who was adamant about rejecting a paper on
grounds that the authors believed were quite unreasonable, but in the third revision the referee did point out
a mistake in a sampling algorithm.  Since publishing a wrong paper is much more damaging in the long run than
being rejected by a given journal, this turned out to be most useful. 

At the other end of the spectrum are sloppy referees who form a strong opinion based on a cursory read, along
with their particular priors about the topic in question.  The result, especially for competitive journals, is
often a rejection based on vague and unrelated comments, which also contributes to an incentive structure that
favours incremental and conventional work. Alternatively, an ``accept'' decision based on shallow refereeing can
allow a poor paper to be published.  Often, however, the system corrects itself, with the discrepancy with the
other reports or the lack of substance in the review being spotted by associate editors or editors.  (We are
not here considering the referees who follow the unethical practice of rejecting a paper and rushing to write
one of their own based on the same idea, which has happened to us only rarely.) 

We also believe that our papers are preemptively improved by refereeing, in that we mostly write better papers
because we know they will be critically evaluated by colleagues prior to publication. We go the extra mile,
chase typos, think more carefully about real examples, and so on, before submitting, because we do not want to
give a negative referee this additional and objective leverage we can ourselves perceive. 

Some authors seem to be outraged when referees demand particular modifications, because they consider that, as
authors, they should have an inalienable right to decide the exact form of their texts. But we are not artists
(who actually rarely obtain this right anyway), we are scientists, and science (in particular hard science,
especially mathematics) works better through consensus on the validity and correctness of the proposed
research.



\section{Wheat from chaff}

\begin{kvothe}
{\em ``The trouble with referees is that they know the rules, but they do not know the game.''}
--- Bill Shankly.
\end{kvothe}

Recognising that scientific review processes have been evolving forever, our current view of the traditional
model is that editors send submitted manuscripts to selected reviewers for comment, and then make a decision
based on these comments and their own judgement. The issues of concern in such a simple system arise from the
arbitrary and often narrow selection of reviewers, the generous and often unreasonable time allowed for
response, the general and often unhelpful guidelines for comments, the opaque manner in which the final
decision is made, and more generally the huge and often wasted investment in time by authors and referees.  In
particular, junior scholars can take their refereeing duties very seriously, writing long and careful reports
on papers which are of such low quality to not be worth the effort.  We agree with Kriegeskorte and Deca (2012) that a
better use of reviewers' time and effort would be to have many reviews of important papers and only zero or one
review for the sorts of minor efforts that fill up our journals.  Conversely, even a very specialised result
can sometimes be useful; in this case it might well merit a post-publication review thread by its user
community, in the same way that more popular restaurants and hotels will receive more reviews on Yelp and
Tripadvisor, giving consumers more information to make their judgements.

Some journals have evolved from this traditional model in order to address these issues. For example, some no
have databases of reviewers from which to more objectively draw subject-specific referees; others demand short
review times; others have formalised the referee process by instituting a list of questions to be answered or
providing careful guidelines about the type of review required; and a small number have adopted the post-modern
(or in fact pre-traditional) practice of the editorial board itself making decisions at regular team meetings. 
 
A strong argument against doing away with referees is the problem of sifting through the chaff. The volume of
research documents published everyday is overwhelming and accelerating, perhaps not so much in statistics but
certainly in biomedical research and engineering. There is a maximum amount of time one can dedicate to looking
at websites, blogs, twitter accounts, and such.  And blog comments have not always delivered the
post-publication quality control that some had hoped.  Commenting on a blog (or, for that matter, a {\em PLoS
One} article) is not a well respected use of time, and there is also the paradox that a comment on a busy blog
might not get noticed amidst all the other comments.  Right now there seem to be very few blogs providing a
useful communal review function (and none of these blogs focus on statistics research).

Even keeping track of new arXiv postings sometimes gets overwhelming. Wasserman writes, ``if you don't check
arXiv for new papers every day, then you are really missing out,'' but our own experience is that it is almost
impossible not to miss out.  Checking the list of the newly posted arXiv in the morning indeed takes less than
a minute, checking potentially interesting papers takes much longer! Even glancing at the 200 or 300 titles at
the end of the month is taking its toll, and more if one starts looking at abstracts and pdf files. 

Without an organised system of reviews, why should anyone bother to comment on poor or wrong, but not
newsworthy, papers?  The result could well be a clutter of mediocre results making it difficult for researchers
who are not well-connected to navigate the field.  We, the authors of the present article, know enough experts
in our research areas that we can often get a quick evaluation of unpublished work. But a student at a middling
university, or one whose advisor is not an expert on statistical computation, or a researcher in biology (say)
who wants to use the latest computational methods, will not generally have the resources that we have in our
social network. The review process does not completely level the playing field---nothing could, given
institutional disparities of resources---but it comes closer to equalising the information available to
differently-equipped teams.

Given the amount of chaff and the connected tendency to choke on it, filtering will be done---somehow or
another. Getting rid of referees and journals in favour of repositories like (the great) arXiv would force us
to rely on other and less well-defined sources for ranking, selecting, and eliminating papers. Again this would
be subject to arbitrariness, subjectivity, bias, variation, randomness, peer pressure, and so on. In addition,
having no prior quality control makes reading a new paper a tremendous chore as one would have to check the
references as well, leading to a type of infinite regress, while forcing one to rely on reputation and peer
opinions.

In fact, one may wonder if it is really possible to go that far in reducing the impact of peer reviewing.  For
many of us, so much depends on our publication record (including jobs, promotions, grants, even salaries in
certain institutions), that very few would be bold enough to stop sending papers to peer-reviewed journals from
their own initiative. Getting rid of peer-reviewed publications would make sense only if the vast majority of
scientists in a given field would agree to do all at once. 
And, since it is not only individuals but also scientific fields that compete for grant money, one could argue
that a simultaneous move from {\em all fields} would be required to ditch peer reviewing, which is of course
even less likely.


Thus, despite the appeal of chucking the journals and starting over, we think a purely uncontrolled system
would be even more unethical than what we currently have, and may be exactly what we would like to avoid in
science and academia. If our profession did start from scratch, we are sure that new institutions would arise
to serve the filtering and reviewing functions, but we would prefer to make that switch more smoothly.   In the
following sections, we make two proposals that constitute a middle ground between what we have now and what
Wasserman is advocating.  The first is a further evolutionary step in the review process, as described above. The
second is a more radical change.

\section{Proposal 1: Post-publication peer review}

\begin{kvothe}
{\em ``The trouble with referees is that they just don't care which side wins.''} --- Tom Canterbury.
\end{kvothe}

In a world where everything (or nearly everything) is published---that is, the world we are currently
approaching---how can the scientific community sift through the mass of results?  It should be possible to use
the effort that is currently going into the peer review system in a more efficient manner.  While writing
dozens of long careful referee reports per year, we realize the futility and effort of creating mini-articles
for such a small audience (the author and the journal editor). It makes much more sense to switch to blogging
about important papers or general points to reach a much wider audience.  And to keep reviews short and to the
point (and available to the readers of the article in question at some point; see below). This notion seems to
be met with reluctance by many of our interlocutors, for whom the secrecy of the reviewing process and the
anonymity of the reviewer appear like sacrosanct principles. To wit, we note the emphasis placed on this
secrecy in a recent update of the {\em Annals of Statistics} guidelines. 

Post-publication peer review could be done in different ways, most simply by adding a comment thread to each
arXiv article (with the caveat of being possibly not read), but more formal approaches are possible.
Kriegeskorte (2009, 2011) recommends ``peer-to-peer editing: authors ask a senior scientist to edit the paper; editor
chooses 3 reviewers and asks them to openly review the paper; editor is named on the paper."

Another, perhaps complementary, approach would be for groups of scholars and academic societies to manage a
filtering service. For example, instead of the American Statistical Association running {\em JASA}, {\em
JABES}, {\em JCGS}, {\em Technometrics}, etc., and maintaining a separate editorial staff for each of those
journals (representing a huge amount of possibly overlapping and hence redundant volunteer service on the part
of editors and referees), it could support filtering services. The editors of each filter would be  expected to
scan the literature and handle submissions (which in this case would be pointers to articles already published
on the web). The editorial boards would have the responsibility to come up with monthly (say) recommended
reading material.  This would require some work, but less than the existing job of producing a journal. The
main concern we see would be to keep the editors focused on solid research rather than getting tabloid-like,
but the latter seems less likely if the process involves simply flagging articles rather than formally and
exclusively publishing them.   The flagging could even be multidimensional, with some papers tagged as
potentially exciting but speculative, and others labelled as solid contributions within an existing paradigm.

Instead of a simple thumbs-up or down, reviewers would have the task of situating each new paper within the
literature.  As journal editors and frequent referees ourselves, we would appreciate the opportunity to prepare
reviews that are directed outward to the potential users of the published articles rather than inward toward
the author and editor.

We suspect that a key step in getting post-publication peer review to work is to transfer the efforts that {\em
would have gone into refereeing} into filtering.  It would be difficult to start up a filter all on its own
without the free labour that comes from referees (who are in turn motivated by a sense of obligation and
scientific community).   The hours in the day are limited, and we foresee a challenge in instilling the same
sense of duty for filtering and post-publication review as is now present in the journal review process.

\section{Proposal 2:  A reviewer commons}

\begin{kvothe}
{\em ``Wheredya get ya ticket, ref---a cereal box?!"}
\end{kvothe}

Just as it's useful to ask why sport referees don't always get it right, we could ask the same of our own
reviewers. What's broken in our system? As with sport, there is a constant proliferation of new arenas of
training and competition and an exponential growth in the community of participants, which has great potential
benefits for science but is daunting for reviewers. However, unlike sport, we don't grow or train our
reviewers. Unlike the training in refereeing that we may have received as a young person, learning to blow a
whistle beside an old hand and watching how others did it, our training in scientific reviewing consisted of
three words: ``Here, review this.''  Instead of the floodlights and open scrutiny to which we subject our
sports referees, scientific reviews are conducted by fluorescent light behind closed doors.  Perhaps it's time
we came out.

As scientists, we continually strive to improve, to find new solutions. Publishing our results is our
lifeblood, so if we don't like what we have at present, we should change it.  However, as scientists, we should
review the present system before discarding it altogether. The referee process is already evolving; so we could
ask what we could we do to accelerate this.


It would help to have open, ongoing peer learning. Sports referees receive continuing training; many
professional societies demand this; we should as well. And we are not solely talking correspondence courses; we
mean the real thing, with floodlights and open scrutiny.  This leads to perhaps our most dramatic suggestion: a
reviewer commons, namely a (virtual) repository for the placement of scientific reviews, open to all. The
advantages of such a commons are many. It would encourage high quality, fair and useful reviews. It would
facilitate acknowledgement of reviewer contributions, benefiting both the journals and authors (since reviews
can be referenced in the manuscript) and the reviewers (since reviews can be referenced in a c.v.~and accessed
by peers). Reviewers would then write not only for the authors but also for the readers, turning their comments
and suggestions into a valuable discussion at the end of the reviewing process, to be added to their
publication list as well. Furthermore, as well as improving quality, this notion of a commons might also help
to reduce the workload of reviewers and editors. For example, until the current practice of not requiring
authors to declare prior submissions of articles is revised, access to previous reviews might help to mitigate
replication of effort by reviewers in dealing with manuscripts doing the rounds of journals and counting on the
geometric distribution to achieve acceptance, sometime, somewhere.

We are not the first to argue that revealing the names of referees, not only to the authors, but also to the
public, would deter referees from being complacent or un-constructively negative. Indeed, it may bring more
explicit recognition in the scientific sense to referees and to their role in publishing better research,
possibly all the way to referees' reports becoming a valued part of their own publication record, as is already
the case for referees for {\em Hydrology and Earth System Sciences}.

A related concern is the increasing focus of some journals on headline-grabbing
articles. This can lead to evaluation of articles on the basis of their popularity rather than their science.
As discussed above, this is against the principles that we laid down for good refereeing practice.
Psychologist Sanjay Srivastava (2011) identifies the problem: 
\begin{quote}
``As long as a journal pursues a strategy of publishing
`wow' studies, it will inevitably contain more unreplicable findings and unsupportable conclusions than equally
rigorous but more 'boring' journals. Ground-breaking will always be higher-risk. And definitive will be the
territory of journals that publish meta-analyses and reviews.''
\end{quote}

Naturally, not all journals are guilty of this charge, and some journals achieve a very strong balance of good
story and high quality science. Moreover, switching to a publish-on-the-arXiv-model will not alone solve the
problem:  it would still be the most dramatic claims that get the most attention.  Post-publication peer review
could help, though.

Again, parallels can be drawn between the concerns expressed above and similar concerns about trends in
professional sport:  increasing corporate control of the game, decreasing public access to the top games (for
example due to restrictive television rights), perception of players as commodities, a focus on spectator
entertainment instead of quality of the game, and so on. While these struggles are widely acknowledged and
debated, by far the majority of the discussion is about improving the system rather than abandoning it
altogether.

\section{Looking forward}

\begin{kvothe}
{\em ``Three cheers for the ref!"}
\end{kvothe}

One cheer for quality, two for fairness, three for excellence. Just as backyard players aspire to higher levels
of play, (true) scientists {\em want} to be reviewed. We want our work to be high quality and accepted by our
peers, and we accept refereeing---and journals---as part of this evaluation excellence. This does not mean that
we must accept poor practice, in terms of quality or ethics, among referees or publishers. Nor does it mean
that having found such faults, we should abolish the system. Indeed, for the self-same reasons of ethics and
quality, it is likely that even if we  did away with scientific refereeing, if we opted instead for a
web-free-for-all, sooner rather than later a system for identifying excellence and equity would emerge. So
instead of evicting, let's try evolving.  Like any good complex system, improvements such as the establishment
of a commons or of society supported post-publication peer review might exhibit similar self-organisation
whereby a more satisfactory process of scientific review evolves of its own accord---or then again, it might
equally implode.

Finally, we have not addressed the problems of lack of replicability raised by empirical researchers in social
science and medicine and which also arise in applied and computational statistics (where, for example, a
published simulation study can be based on a complex tangle of code that can sometimes not be reproduced by the
original researchers, let alone outsiders).  Just as biomedical journals are moving toward registration of
protocols and data, statistics researchers might soon be expected to produce replicable papers with R/Matlab/C
code, data, and random seeds for all simulations.  Perhaps replication will be part of the screening and
vetting process:  the aggregators that replace traditional journals can note a paper as having its replication
materials or not, and perhaps the top aggregators (for example, the replacements for {\em JASA} and {\em JRSS}) will
require replicability to get their stamp of approval.

\section*{Acknowledgements} 
We thank in advance (!) the referees of this article for their useful comments, the journal editors for
accepting the manuscript, and the publisher for staying in business in order to publish our work.  

The first (NC) and last authors' (CPR) research is partly supported by the Agence Nationale de la Recherche
(ANR, 212, rue de Bercy 75012 Paris) through the 2008--2012 grant ANR-08-BLAN-0237 ``Big'MC.''  The second
author (AG) thanks the U.S. National Science Foundation for partial support of this work.  The third author
(KM) acknowledges the generosity of CEREMADE and CREST as hosts while this article was written.  

\renewcommand{\bibsection}{\section*{References}}


\end{document}